\documentclass[10pt,journal,twoside]{IEEEtran}
\IEEEoverridecommandlockouts
\usepackage{cite}
\usepackage{amsmath,amssymb,amsfonts}
\usepackage{algorithmic}
\usepackage{graphicx}
\usepackage{textcomp}
\usepackage{xcolor}
\usepackage{hyperref}
\usepackage{CJKutf8}
\usepackage{algorithm}
\usepackage{setspace}
\usepackage{array}
\usepackage[caption=false,font=normalsize,labelfont=sf,textfont=sf]{subfig}
\usepackage{textcomp}
\usepackage{stfloats}
\usepackage{url}
\usepackage{verbatim}
\def\BibTeX{{\rm B\kern-.05em{\sc i\kern-.025em b}\kern-.08em
		T\kern-.1667em\lower.7ex\hbox{E}\kern-.125emX}}
\begin{document}
	\begin{CJK}{UTF8}{gbsn}
		\vspace{-0.6cm}
		\title{\huge Rotatable Antenna-Enabled Secure Wireless Communication}
		\author{Liang Dai, Beixiong Zheng,~\IEEEmembership{Senior Member,~IEEE,} Qingjie Wu, Changsheng You,~\IEEEmembership{Member,~IEEE,}\\Robert Schober,~\IEEEmembership{Fellow,~IEEE,} and Rui Zhang,~\IEEEmembership{Fellow,~IEEE}
		\vspace{-0.3cm}
	    \thanks{L. Dai, B. Zheng, and Q. Wu are with the School of Microelectronics, South China University of Technology, Guangzhou 511442, China (e-mails: 202321061996@mail.scut.edu.cn; bxzheng@scut.edu.cn; miqjwu@mail.scut.edu.cn).\par
	    C. You is with the Department of Electronic and Electrical Engineering, Southern University of Science and Technology, Shenzhen 518055, China and also with the Shenzhen Key Laboratory of Optoelectronics and Intelligent Sensing, Shenzhen 518055, China (e-mail: youcs@sustech.edu.cn).\par
	    Robert Schober is with the Institute for Digital Communications, Friedrich-Alexander-University Erlangen-Nürnberg (FAU), 91054 Erlangen, Germany (e-mail: robert.schober@fau.de).\par
	    R. Zhang is with School of Science and Engineering, Shenzhen Research Institute of Big Data, The Chinese University of Hong Kong, Shenzhen 518172, China (e-mail: rzhang@cuhk.edu.cn), and also with the Department of Electrical and Computer Engineering, National University of Singapore, Singapore 117583 (e-mail: elezhang@nus.edu.sg).
	    }
		\vspace{-0.7cm}
		}
		\maketitle
        
		\begin{abstract}
		Rotatable antenna (RA) is a promising technology that exploits new spatial degree-of-freedom (DoF) to improve wireless communication and sensing performance. In this letter, we investigate an RA-enabled secure communication system where confidential information is transmitted from an RA-based access point (AP) to a single-antenna legitimate user in the presence of multiple eavesdroppers. We aim to maximize the secrecy rate by jointly optimizing the transmit beamforming and the deflection angles of all RAs at the AP. Accordingly, we propose an efficient alternating optimization (AO) algorithm to obtain a high-quality suboptimal solution in an iterative manner, where the generalized Rayleigh quotient-based beamforming is applied and the RAs' deflection angles are optimized by the successive convex approximation (SCA) technique. Our simulation results show that the proposed RA-enabled secure communication system achieves a significantly higher secrecy rate as compared to various benchmark schemes.
		\end{abstract}
		
		
		\section{Introduction}
        With the rapid development of global information and communication technology (ICT), the forthcoming wireless communication network is envisioned to enable intelligent connectivity for a drastically increasing number of devices and users, leading to the circulation of massive data. However, due to the broadcast nature of wireless signals, the communication links are highly vulnerable to interception by eavesdroppers, resulting in sensitive data leakage. To mitigate such security threats, various technologies have been developed, such as high-layer encryption techniques\cite{01} and physical layer security (PLS) technologies\cite{0}. Notably, PLS technologies, which enhance communication security by leveraging physical transmission characteristics, have attracted significant research attention. Specifically, various PLS techniques have been developed to improve secrecy rate, including artificial noise (AN)-based jamming \cite{1,2,3}, multi-antenna beamforming\cite{4,5}, and precoding design\cite{6}. Nevertheless, these technologies are inherently constrained by the conventional antenna architecture with fixed positions and orientations of antennas once they are deployed. This limitation hinders the full exploitation of spatial channel variation at wireless transceiver and compromises secrecy performance, thus motivating the need for more flexible and adaptive PLS designs.\par
       \begin{figure}[t]
        	\centering
        	\vspace{-0.2cm}
        	\includegraphics[width=2.4in]{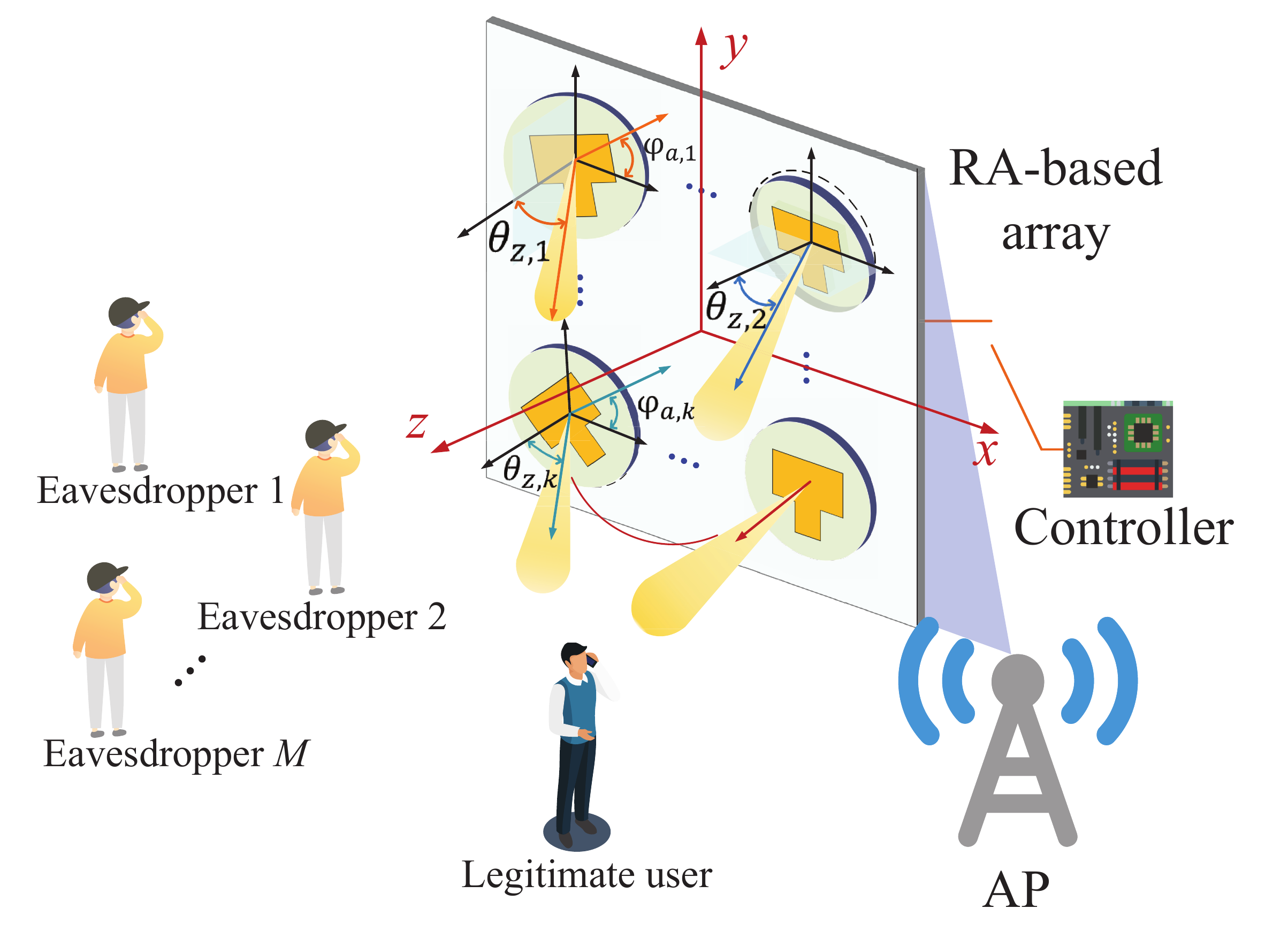}
        	\setlength{\abovecaptionskip}{-4pt}
        	\vspace{-0.2cm}
        	\caption{An RA-enabled secure communication system.}
        	\label{fig1}
        	\vspace{-0.6cm}
        \end{figure}
        Recently, rotatable antenna (RA) has been proposed as a promising technology to exploit new spatial degree-of-freedom (DoF) by flexibly adjusting its three-dimensional (3D) orientation/boresight direction in response to wireless channel variations\cite{7,8}. In particular, RA can be considered as a simplified implementation of the six-dimensional movable antenna (6DMA) \cite{11} with antenna rotation only, which significantly reduces hardware complexity with a highly compact design. Specifically, RA-based arrays can change their directional gain patterns by independently adjusting the deflection angles of all antenna elements, thereby collaboratively enhancing the overall array gain in desired directions while suppressing the radiation power in undesired directions. Some initial studies have demonstrated the great potential of RA-enabled communication systems\cite{7,8,9,10}. In particular, the authors in \cite{7,8} established the fundamental system model and channel characterization of RA systems for the first time. Furthermore, the authors in \cite{8} provided theoretical analyses that validated the performance gains enabled by RAs. The authors in \cite{9} investigated a new RA-assisted integrated sensing and communications (ISAC) system and revealed the improvement in terms of weighted communication and sensing performance. Owing to its ability to flexibly adjust the 3D orientation/boresight direction of each antenna, RA can selectively enhance and/or suppress signal power in specific spatial directions. This capability makes RA particularly appealing for PLS design. Nevertheless, despite these promising capabilities, the potential of RA to facilitate secure communication remains unexplored, thus deserving a dedicated study as pursued in this letter.\par
        Motivated by the above, we consider an RA-enabled PLS system as shown in Fig. \ref{fig1}, where an RA-based array is deployed at the access point (AP) to serve a legitimate user in the presence of multiple eavesdroppers. To enhance the communication quality of the legitimate user while preventing information leakage to the eavesdroppers, we formulate an optimization problem for maximization of the secrecy rate by jointly optimizing the transmit beamforming of the AP and the deflection angles of all its RAs, subject to constraints on the RAs' deflection angles and the transmit power. As the formulated problem is non-convex, we propose an alternating optimization (AO) algorithm to obtain a high-quality suboptimal solution by using the successive convex approximation (SCA) and the generalized Rayleigh quotient techniques. Our simulation results show that RAs can significantly enhance the secrecy rate and outperform other benchmark schemes, such as conventional fixed antennas-based system.\par
		\section{System Model and Problem Formulation}
		As illustrated in Fig. \ref{fig1}, we consider a PLS system where the AP intends to transmit confidential information to a legitimate user in the presence of $M \geq 1$ eavesdroppers. The AP is equipped with a uniform planar array (UPA) consisting of $K$ directional RAs, while the legitimate user and each eavesdropper are equipped with a single isotropic and fixed antenna. We assume that the UPA at the AP is placed on the $x$-$y$ plane of a 3D Cartesian coordinate system. Additionally, $\mathbf{q}_m \in \mathbb{R}^{3 \times 1}$, $\forall m \in \{0,1,\dots,M\}$, represent the reference positions of the legitimate user and the eavesdroppers, where $\mathbf{q}_0$ and $\mathbf{q}_m$, $m \geq 1$, are the reference positions of the legitimate user and the $m$-th eavesdropper, respectively. Let $\mathbf{w}_k \in \mathbb{R}^{3 \times 1}$ denotes the reference position of the $k$-th RA. The deflection angle vector of the $k$-th RA is defined as $\boldsymbol{\theta}_k = \begin{bmatrix}\theta_{{z},k},\varphi_{{a},k} \end{bmatrix}^T$, $1\leq {k} \leq {K}$, where the zenith angle $\theta_{{z},k}$ represents the angle between the boresight direction of the $k$-th RA and the $z$-axis, and the azimuth angle $\varphi_{{a},k}$ denotes the angle between the projection of the boresight direction of the $k$-th RA onto the $x$-$y$ plane and the positive $x$-axis. Note that the 3D orientation/boresight direction of each RA can be mechanically and/or electrically adjusted by a common smart controller\cite{7,8}. Specifically, the pointing vector of each RA is defined as $\mathbf{f}(\boldsymbol{\theta}_k) = [ \sin\theta_{{z},k} \cos\varphi_{{a},k}, \sin\theta_{{z},k}  \sin\varphi_{{a},k}, \cos\theta_{{z},k} ]^T$ with $\|\mathbf{f}(\boldsymbol{\theta}_k)\| = 1$ for normalization, where $\|\cdot\|$ denotes the Euclidean norm. The effective antenna gain for each RA satisfies the following generic directional gain pattern\cite{8},
		\vspace{-0.15cm}
		\begin{equation}
			\resizebox{0.8\hsize}{!}{$
			G_e(\epsilon, \varphi) = 
			\begin{cases} 
				G_0 \cos^{2p}(\epsilon), & \epsilon \in [0, \frac{\pi}{2}), \varphi \in [0, 2\pi) \\
				0, & \text{otherwise},
			\end{cases}$}
		\end{equation}
		where \( (\epsilon, \varphi) \) is a pair of incident angles corresponding to any direction with respect to the RA's boresight direction, $G_0 = 2(2p +1)$ denotes the maximum gain in the boresight direction to meet the law of power conservation, and $p$ denotes the directivity factor that characterizes the beamwidth of the antenna’s main lobe. Due to the boresight symmetry of the antenna gain pattern, its directional gain depends solely on angle $\epsilon$.
		Accordingly, the $k$-th RA's directional gains in the direction of the legitimate user and the $m$-th eavesdropper are given by $G_{m,k} = G_0 \cos^{2p}(\epsilon_{m,k})$, where $\cos(\epsilon_{m,k}) \triangleq  \mathbf{f}^T(\boldsymbol{\theta}_k) \vec{\mathbf{q}}_{m,k}$ is the projection between the direction vector of the ${m}$-th eavesdropper (or legitimate user with $m=0$) $\vec{\mathbf{q}}_{m,k} \triangleq \frac{\mathbf{q}_m - \mathbf{w}_k}{\|\mathbf{q}_m - \mathbf{w}_k\|} $ and the pointing vector of the ${k}$-th RA. \par
		Assuming that all the channels in the considered system experience quasi-static flat-fading, the channel from the ${k}$-th RA to the ${m}$-th eavesdropper (or legitimate user with $m=0$) can be modeled as
			\vspace{-0.1555cm}
		\begin{equation}
			h_{m,k}(\boldsymbol{\theta}_k)= \sqrt{L(d_{m,k}) G_{m,k}} {g}_{m,k}. \label{h2}
		\end{equation}
		Here, $L(d_{m,k})$ denotes the large-scale channel power gain characterizing the distance-dependent path loss and shadowing effect, which can be modeled as
			\vspace{-0.15cm}
		\begin{equation}
			L(d_{m,k}) = {\zeta_0 \left(d_0/d_{m,k}\right)^{\alpha_m}},
		\end{equation} 
		where $\zeta_0$ represents the channel power gain at the reference distance of $d_0 = 1$ meter (m), $\alpha_m$ represents the path loss exponent, and $d_{m,k}$ denotes the distance from the ${k}$-th RA to the ${m}$-th eavesdropper (or legitimate user with $m=0$).\par
		Furthermore, ${g}_{m,k}$ is the small-scale channel fading coefficient, which is modeled by independent Rician distribution as follows, with Rician factor $K_m$.\\
			\vspace{-0.15cm}
		\begin{equation}
			{g}_{m,k} = \sqrt{\frac{K_m}{K_m+1}}\bar{g}_{{m,k}} + \sqrt{\frac{1}{K_m+1}}\tilde{g}_{{m,k}},
		\end{equation}
		where \(\bar{g}_{{m,k}}=e^{-j\frac{2\pi}{\lambda}d_{{m,k}}}\) denotes the line-of-sight (LoS) channel component, and $\tilde{g}_{{m,k}} \sim \mathcal{CN}(0, 1)$ denotes the non-LoS (NLoS) channel component characterized by Rayleigh fading. Therefore, the overall channel from AP to the ${m}$-th eavesdropper (or legitimate user with $m=0$) is defined as 
			\vspace{-0.1cm}
		\begin{equation}
			 \mathbf{h}_{m}(\boldsymbol{\Theta}) = [h_{m,1}(\boldsymbol{\theta}_1), 
			h_{m,2}(\boldsymbol{\theta}_2), ..., h_{m,K}(\boldsymbol{\theta}_K)]^T,
		\end{equation}
		where $\boldsymbol{\Theta} \triangleq [
		\boldsymbol{\theta}_1, \boldsymbol{\theta}_2, \ldots, \boldsymbol{\theta}_{K}]\in \mathbb{R}^{2 \times {K}}$ denotes the deflection angle matrix.
		For the purpose of exposition, we assume that the global channel state information (CSI) of all the above channels has been acquired/estimated at the AP for the joint design of all RA deflection angles and transmit beamforming weights. This assumption holds in scenarios where eavesdroppers are active but untrusted participants in the system. \par
		The AP transmits confidential information $s$ with zero mean and unit variance to the legitimate user via beamforming and orientation adjustment. The beamforming vector is denoted by $\mathbf{v}^H\in \mathbb{C}^{1 \times {K}}$, which satisfies the following power constraint.
			\vspace{-0.15cm}
		\begin{equation}
			\|\mathbf{v}\|^2 \leq P_\mathrm{{AP}},
		\end{equation}
		where ${P}_\mathrm{{AP}}$ is the maximum transmit power of the AP and the superscript $(\cdot)^H$ represents the Hermitian transpose. Thus, the received signals at the legitimate user and the ${m}$-th eavesdropper are respectively expressed as
			\vspace{-0.15cm}
		\begin{align}
			{y_{0}} &= \mathbf{v}^H \mathbf{h}_{0}(\boldsymbol{\Theta}) s +  n_{0}, \\
			{y_{{m}}} &= \mathbf{v}^H \mathbf{h}_{m}(\boldsymbol{\Theta}) s +  n_{m},
		\end{align}		
	   where $n_{0}$ and $n_{m}$ denote the additive white Gaussian noise (AWGN) at the legitimate user and the ${m}$-th eavesdropper, with zero-mean and variances $\sigma_\text{r}^2$ and $\sigma_\text{e}^2$, respectively. Considering the worst case where ${M}$ eavesdroppers collaborate to decode the intercepted confidential information, the secrecy rate in bits/second/Hz (bps/Hz) is given by
	   	\vspace{-0.1cm}
		\begin{equation}
		R_{\mathrm{sec}} = \left[ R_\mathrm{u} - R_\mathrm{e} \right]^+,
		\end{equation}
		where $[x]^+ \triangleq \max\{x, 0\}$, and
			\vspace{-0.1cm}
	\begin{align}
					&R_\mathrm{u} = \log_2 \left( 1 + \frac{\left| \mathbf{v}^H \mathbf{h}_{0}(\boldsymbol{\Theta}) \right|^2}{\sigma_\mathrm{r}^2}\right),\\
					&R_\mathrm{e} = \log_2 \left( 1 + \frac{\sum_{m=1}^M \left| \mathbf{v}^H \mathbf{h}_{m}(\boldsymbol{\Theta}) \right|^2 }{\sigma_\mathrm{e}^2}\right),
				\end{align}
		are the achievable rates of the legitimate link and the eavesdropper links, respectively.\par
		Our goal is to maximize the secrecy rate $R_{\mathrm{sec}}$ by jointly optimizing the AP transmit beamforming vector $\mathbf{v}$ and the deflection angle matrix $\boldsymbol{\Theta}$ of all RAs, subject to constraints on zenith angle and the transmit power. Thus, the optimization problem can be expressed as
		\vspace{-0.15cm}
		\begin{align}
			(\text{P1}) \quad \max_{\mathbf{v}, \boldsymbol{\Theta}}&\quad R_{\mathrm{sec}} \tag{12a}\\ 
			\text{s.t.} &\quad 0 \leq \theta_{z,k} \leq \theta_{\max},\; \forall k, \tag{12b} \\
		    &\quad \| \mathbf{v} \|^2 \leq P_\mathrm{AP}, \tag{12c}
	   \end{align}
	where $\theta_{\max}$ represents the maximum zenith angle that each RA is allowed to adjust. Note that in the objective function of (P1), we can omit the operator $[\cdot]^+$ without loss of optimality, since the optimal value of this problem must be non-negative. However, the objective function of (P1) is non-concave and the two optimization variables $\boldsymbol{\Theta}$ and $\mathbf{v}$ are coupled with each other. Therefore, an AO algorithm is proposed to iteratively optimize the transmit beamforming and deflection angles.
		\section{Proposed Algorithm for Problem (P1)}
		In this section, we develop an AO algorithm to solve problem (P1). Specifically, the generalized Rayleigh quotient technique is used for designing the
		beamforming vector, while the SCA technique is used to design the deflection angles.
		\vspace{-0.3cm}
		\subsection{Transmit Beamforming Optimization}
		For a given deflection angle matrix $\boldsymbol{\Theta}$, by defining 
		\vspace{-0.1cm}
		\begin{align}
		\setcounter{equation}{12} 
		\mathbf{A} &= \frac{\mathbf{h}_{0}(\boldsymbol{\Theta}) \mathbf{h}_{0}(\boldsymbol{\Theta})^H}{\sigma_\mathrm{r}^2},\\
		\mathbf{B} &= \frac{\sum_{{m}=1}^{M} \mathbf{h}_{m}(\boldsymbol{\Theta}) \mathbf{h}_{m}(\boldsymbol{\Theta})^H}{\sigma_\mathrm{e}^2}, 
     	\end{align}
		problem (P1) can be reformulated as
		\vspace{-0.15cm}
		\begin{align}
		(\text{P2})\quad \max_{\mathbf{v}} &\quad  \log_2 \left(\frac{1 + \mathbf{v}^H \mathbf{A} \mathbf{v}}{1 + \mathbf{v}^H \mathbf{B} \mathbf{v}}\right) \tag{15a} \\
		\text{s.t.} &\quad\|\mathbf{v}\|^2 \leq P_\mathrm{{AP}}. \tag{15b}
		\end{align}
	   According to the generalized Rayleigh quotient technique, the optimal solution to (P2) is \cite{4}
	   \setcounter{equation}{15} 
	   \vspace{-0.1cm}
		\begin{equation}
		\mathbf{v} = \sqrt{P_\mathrm{{AP}}} \mathbf{o}_{\max}, \label{v_opt}
        \end{equation}
	where $\mathbf{o}_{\text{max}}$ is the normalized eigenvector corresponding to the largest eigenvalue of matrix $\left( \mathbf{B} + \frac{1}{P_\mathrm{{AP}}} \mathbf{I}_{K} \right)^{-1}\left( \mathbf{A} + \frac{1}{P_\mathrm{{AP}}} \mathbf{I}_{K} \right) $, with $\mathbf{I}_{K}$ being the ${K} \times {K}$ identity matrix and $(\cdot)^{-1}$ being the matrix inversion.
	\subsection{Deflection Angle Optimization}
	For a given transmit beamforming matrix $\mathbf{v}$, problem (P1) can be reformulated as
	\vspace{-0.2cm}
	\begin{align}
		(\text{P3}) \quad \max_{\boldsymbol{\Theta}}&\quad R_{\text{sec}} \tag{17a}\\ 
		\text{s.t.} &\quad 0 \leq \theta_{{z,k}} \leq \theta_{\max},\; \forall{k}. \tag{17b} 
	\end{align}
	It is obvious that the deflection angles mainly affect the channel power gain through the projection $\cos({\epsilon}_{{m,k}})$. For the convenience of our subsequent derivations, an auxiliary variable $\mathbf{f}_{k} \in \mathbb{R}^{3 \times 1}$ is introduced to equivalently represent the effect of deflection angle vector $\boldsymbol{\theta}_{k}$ on the pointing vector of the ${k}$-th RA, i.e., $\mathbf{f}_{k} \triangleq \mathbf{f}(\mathbf{\boldsymbol{\theta}}_{k})$. In particular, the channel from the ${k}$-th RA to the ${m}$-th eavesdropper/legitimate user is reformulated as 
	\vspace{-0.15cm}
	\begin{align}
		\setcounter{equation}{17} 
		\tilde{h}_{m,k}(\mathbf{f}_{k})& = \tilde{\beta}_{m,k} \left(\mathbf{f}_{k}^T \vec{\mathbf{u}}_{m,k} \right)^p,
	\end{align}
	 where $\tilde{\beta}_{{m,k}} \triangleq \sqrt{L(d_{{m,k}}) G_0} g_{{m,k}}$. Defining $\mathbf{F} \triangleq [
	\mathbf{f}_1, \mathbf{f}_2, \ldots, \mathbf{f}_{K}]$, the channel vectors $\mathbf{h}_{m}(\boldsymbol{\Theta})$ can be expressed as $\mathbf{h}_{m}(\mathbf{F}) \triangleq [\tilde{h}_{m,1}(\mathbf{f}_1), \tilde{h}_{m,2}(\mathbf{f}_2), \ldots, \tilde{h}_{m,K}(\mathbf{f}_{K})]^T$. Thus, problem (P3) can be reformulated as
	\vspace{-0.15cm}
    \begin{align}
		(\text{P4}) \quad \max_{\mathbf{F}} \quad &\log_2 \left(1 + \frac{\left| \mathbf{v}^H \mathbf{h}_{0}(\mathbf{F}) \right|^2}{\sigma_\mathrm{r}^2} \right) \quad \tag{19a}\\&- \log_2 \left(1 + \frac{\sum_{{m=1}}^{M} \left|\mathbf{v}^H \mathbf{h}_{m}(\mathbf{F}) \right|^2}{\sigma_\mathrm{e}^2}\right) \nonumber\\ 
    	 \text{s.t.} &\quad \mathbf{f}_{k}^T \mathbf{e}_3 \geq \cos(\theta_{\max}), \, \; \forall {k}, \tag{19b}\\
    	&\quad  \|\mathbf{f}_{k}\| = 1, \, \; \forall {k}, \tag{19c}
    \end{align}
    where $\mathbf{e}_3 = [0,0,1]^T$, constraint (19b) is equivalent to (17b), and constraint (19c) ensures that $\mathbf{f}_{k}$ is a unit vector. Problem $(\text{P4})$ is still a non-convex optimization problem, which is challenging to solve directly. To tackle this problem, the SCA technique is used to approximate problem $(\text{P4})$ as a convex problem and thereby obtain a local optimal solution. Without loss of generality, we present the procedure for the $(i+1)
    $-th iteration in the following. $\mathbf{F}^{(i)}$ and $R_{\mathrm{sec}}^{(i)}$ denote the solutions of $\mathbf{F}$ and $R_{\mathrm{sec}}$ obtained in the $i$-th iteration. By utilizing the first-order Taylor expansions at ${\mathbf{f}_{k}^{(i)}}$, the expressions $\log_2 \left(1 + \frac{\left| \mathbf{v}^H \mathbf{h}_{0}(\mathbf{F}) \right|^2}{\sigma_\mathrm{r}^2} \right)$ and $\log_2 \left(1 +\frac{\sum_{{m=1}}^{M} \left|\mathbf{v}^H \mathbf{h}_{m}(\mathbf{F}) \right|^2}{\sigma_\mathrm{e}^2}\right)$ in problem $(\text{P4})$ can be linearized  as $ \Lambda_{r}^{(i+1)}(\mathbf{F})$ and  $\Omega_{M}^{(i+1)}(\mathbf{F})$, as shown at the top of the next page, where 
    \vspace{-0.1cm}
    \begin{equation}
    	  \setcounter{equation}{22} 
    		\mathbf{\bar{h}}_{{m,k}}' = \frac{\partial \tilde{h}_{{m,k}}\left(\mathbf{f}_{k}^{(i)}\right)}{\partial \mathbf{f}_{k}^{(i)}} = \left(\tilde{\beta}_{{m,k}} p \left( \left( \mathbf{f}_{k}^{(i)} \right)^T \vec{\mathbf{u}}_{{m,k}} \right)^{p-1}\right) \vec{\mathbf{u}}_{{m,k}},
    \end{equation}
    and $(\cdot)^*$ represents the conjugation. Thus, in the $(i+1)$-th iteration, problem $\text{(P4)}$ can be approximated by the following problem.
    \vspace{-0.1cm}
    	\begin{align}
    		(\text{P5}) \quad \max_{\mathbf{F}}&\quad \Lambda_{{r}}^{(i+1)}(\mathbf{F}) - \Omega_{{M}}^{(i+1)}(\mathbf{F}) \tag{23} \\ 
    		\text{s.t.}&\quad \mathrm{(19b)}, \mathrm{(19c)}. \nonumber 		   		
    	\end{align}\par
    \begin{figure*}[h] 
    	\vspace{-0.7cm}
    	\begin{equation}
    		\setcounter{equation}{20} 
    		\Lambda_{{r}}^{(i+1)}(\mathbf{F}) \triangleq \log_2 \left( 1 + \frac{\left| \mathbf{v}^H \mathbf{h}_{0}(\mathbf{F}^{(i)}) \right|^2}{\sigma_\mathrm{r}^2} \right) + \frac{1}{\ln 2} \frac{\frac{1}{{\sigma_\mathrm{r}^2}} \Re \left\{ \left( \mathbf{v}^H \mathbf{h}_{0}(\mathbf{F}^{(i)}) \right)^*\sum_{{k=1}}^{K} {v}_k^* (\mathbf{\bar{{h}}}_{{0,k}}')^T (\mathbf{f}_k - \mathbf{f}_k^{(i)}) \right\}}{1 + \frac{\left| \mathbf{v}^H \mathbf{h}_{0}(\mathbf{F}^{(i)}) \right|^2}{\sigma_\mathrm{r}^2}},
    	\end{equation}
    		\vspace{-0.15cm}
    	\begin{equation}
    		\Omega_{M}^{(i+1)}(\mathbf{F}) \triangleq \log_2 \left( 1 + \frac{\sum_{{m=1}}^{M} \left| \mathbf{v}^H \mathbf{h}_{m}(\mathbf{F}^{(i)}) \right|^2}{\sigma_\mathrm{e}^2} \right) + \frac{1}{\ln 2} \frac{\sum_{{m=1}}^{M} \frac{1}{{\sigma_\mathrm{e}^2}} \Re \left\{ \left( \mathbf{v}^H \mathbf{h}_{m}(\mathbf{F}^{(i)}) \right)^* \sum_{{k=1}}^{K} {v}_k^* 	(\mathbf{\bar{{h}}}_{{m,k}}')^T (\mathbf{f}_k - \mathbf{f}_k^{(i)}) \right\}}{1 + \frac{\sum_{{m=1}}^{M} \left| \mathbf{v}^H \mathbf{h}_{m}(\mathbf{F}^{(i)}) \right|^2}{\sigma_\mathrm{e}^2}}.
    	\end{equation}
    	\hrulefill
    	\vspace{-0.2cm}
    \end{figure*}
    However, problem $\text{(P5)}$ is still non-convex due to the unit constraint for $\mathbf{f}_k$. For convenience, we relax the equality constraint (19c) as $\|\mathbf{f}_k\| \leq 1$, which yields the following problem.
    \vspace{-0.1cm}
    	\begin{align}
    		(\text{P6})\quad \max_{\mathbf{F}} &\quad\Lambda_{{r}}^{(i+1)}(\mathbf{F}) - \Omega_{{M}}^{(i+1)}(\mathbf{F}) \tag{24a} \\ 
    		\text{s.t.} &\quad \mathbf{f}_{k}^T \mathbf{e}_3 \geq \cos(\theta_{\max}), \, \; \forall {k}, \tag{24b}  \\
    		&\quad \|\mathbf{f}_{k}\| \leq 1, \; \forall {k}. \tag{24c}		
    	\end{align}
    	Problem $\text{(P6)}$ is a convex optimization problem, which can be solved by the CVX solver. Note that the optimal value obtained from problem (P6) provides an upper bound for problem (P5) due to the relaxation of equality constraint (24c). After obtaining the optimal solutions of $\mathbf{v}$ and $\mathbf{F}$ via Algorithm \ref{alg1}, we normalize each pointing vector $\mathbf{f}_k$ in $\mathbf{F}$ to  satisfy constraint (19c), i.e., $\mathbf{f}_k^{\star} = \frac{\mathbf{f}_k}{\|\mathbf{f}_k\|}$. Furthermore, since the original variable of problem (P1) is the RA deflection angle matrix $\mathbf{\Theta}$, an additional step is required to transform the optimized pointing vector into the desired deflection angles\cite{8}.
    	\subsection{Overall Algorithm}
    	   \begin{algorithm}	  
    		\caption{Proposed AO Algorithm for Solving (P1).}
    		\label{alg1}
    		\begin{algorithmic}[t]
    			\STATE \textbf{Initialize:} Pointing vector $\mathbf{F}^{(0)} = \begin{bmatrix}
    				\mathbf{e}_3, \cdots, \mathbf{e}_3
    			\end{bmatrix}_{3 \times K}$, threshold $\varepsilon = 10^{-8}$, and $i = 0$.
    			\REPEAT    			
    			\STATE Given $\mathbf{F}^{(i)}$, calculate $\mathbf{v}^{(i+1)}$ according to (16).
    			\STATE Given $\mathbf{v}^{(i+1)}$ and $\mathbf{F}^{(i)}$, obtain $\mathbf{F}^{(i+1)}$ by solving problem (P6).
    			\STATE Update $i = i + 1$.
    			\UNTIL{$\left|\frac{R_\mathrm{{sec}}^{(i+1)} - R_\mathrm{{sec}}^{(i)}}{R_\mathrm{{sec}}^{(i)}}\right| \leq \varepsilon$.}
    			\STATE \textbf{Output:} $\mathbf{v} = \mathbf{v}^{(i)}$ and $\mathbf{F} = \mathbf{F}^{(i)}$.
    		\end{algorithmic}
    	\end{algorithm}
    	The overall AO algorithm to solve (P1) is summarized in Algorithm 1. Since the optimal objective function $R_{\text{sec}}$ is non-decreasing over the iterations and upper-bounded, Algorithm 1 is guaranteed to converge. The complexity order of problem (P2) is $\mathcal{O}({K}^3)$, while that of problem (P6) is $\mathcal{O}({K}^{3.5}\ln(1/\varepsilon))$. Therefore, the overall complexity of solving (P1) is $\mathcal{O}(L\cdot ({K}^3 +  {K}^{3.5}\ln(1/\varepsilon)))$, where $L$ is the number of iterations required for convergence. 

	\section{Simulation Results}
	In this section, we present simulation results to verify the performance of our proposed RA-enabled secure wireless communication system with the proposed AO algorithm. In
	the following simulations, we assume the system operates at a frequency of 2.4 GHz, corresponding to a wavelength of $\lambda = 0.125$ m. The AP is equipped with a UPA consisting of ${K} = {{k_x}} {{k_y}}$ RAs, where ${{k_x}}$ and ${{k_y}}$ represent the numbers of RAs along the $x$- and $y$-axes, respectively. The array is centered at the origin, with an inter-antenna spacing of $d = \frac{\lambda}{2}$, directivity factor $p = 4$, and the maximum zenith angle $\theta_{\max} = \frac{\pi}{6}$. Specifically, unless otherwise stated, we set ${{k_x}} = {{k_y}} = {k} = 4$, $\zeta_0 = -30 $ dB, and $\alpha_{m} = 3$. The Rician factor is set as ${K_m} = 1$, and the noise powers are set to $\sigma_\mathrm{r}^2 = \sigma_\mathrm{e}^2 = -60$ dBm. We assume that the legitimate user is located at $[r_0\cos\phi_0$, 0, $r_0\sin\phi_0]$, and ${M} = 3$ eavesdroppers are located at $[r_1\cos\phi_1$, 0, $r_1\sin\phi_1]$,  $[r_1\cos\phi_2$, 0, $r_1\sin\phi_2]$, and $[r_1\cos\phi_3$, 0, $r_1\sin\phi_3]$, respectively, where $r_0 = 50$ m and $r_1 = 70$ m are the distances from the origin to the legitimate user and the eavesdroppers, respectively. Moreover, $\phi_i \in [0, \pi], \forall i\in\{0, 1,\dots, {M\}}$, represent the angles between the $x$-axis and the directions of both the legitimate user and the eavesdroppers. We set $\phi_0 = \frac{\pi}{3}$, $\phi_1 = \frac{5\pi}{12}$, $\phi_2 = \frac{2\pi}{3}$, and $\phi_3 = \frac{\pi}{6}$. All simulation results are averaged over 200 random channel fading realizations to ensure accurate statistical averaging.\par
	 \begin{figure}[t]
		\vspace{-0.5cm}
		\centering
		\includegraphics[width=2.2in]{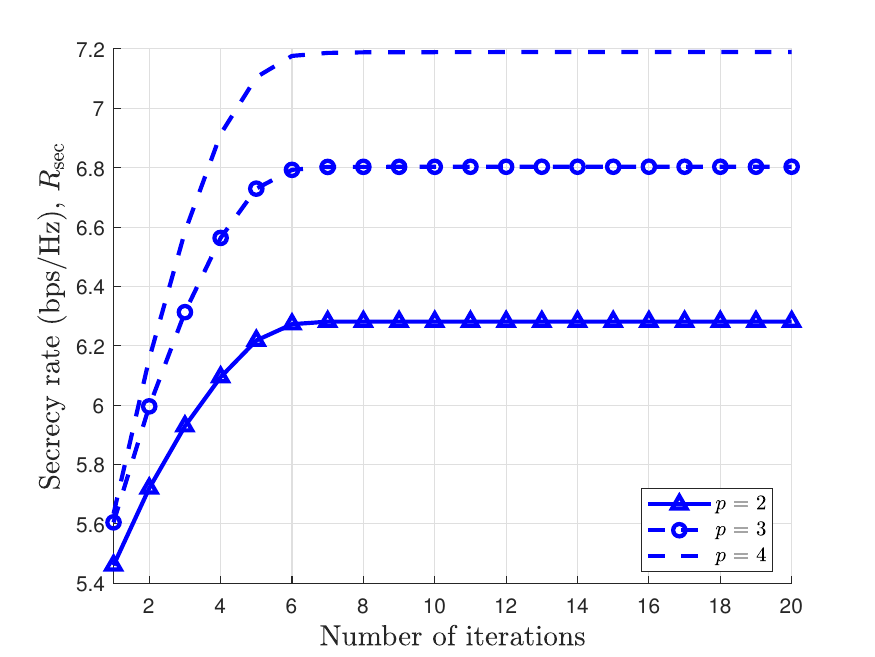}
		\setlength{\abovecaptionskip}{-4pt}
		\vspace{0cm}
		\caption{Convergence behavior of the proposed AO algorithm.}
		\label{fig6}
		\includegraphics[width=2.2in]{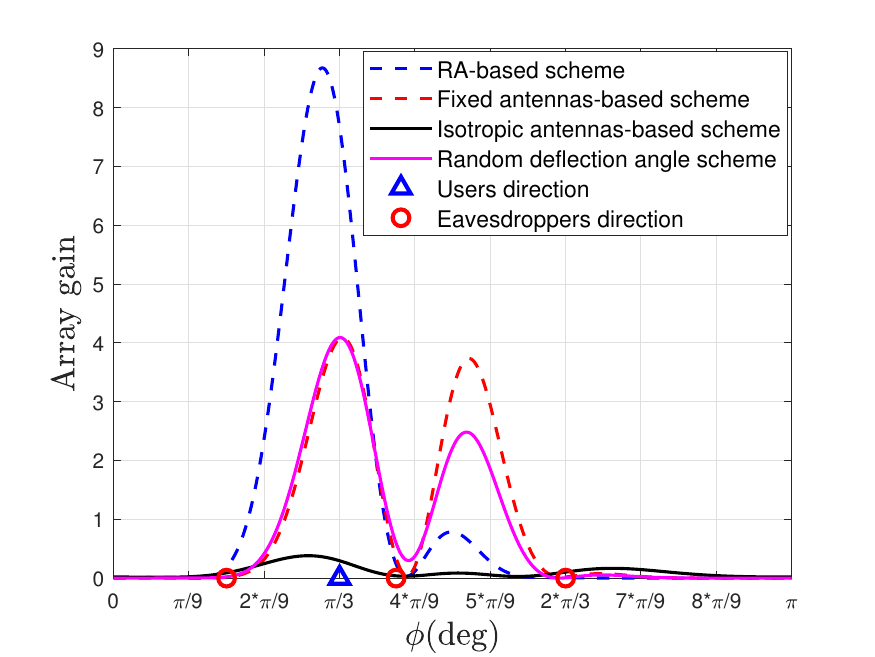}
		\setlength{\abovecaptionskip}{-4pt}
		\vspace{0cm}
		\caption{Comparison of array gain patterns for different schemes.}
		\label{fig7}
		\vspace{-0.6cm}
	\end{figure}
		\begin{figure*}
		\setlength{\abovecaptionskip}{-5pt}
		\setlength{\belowcaptionskip}{-10pt}
		\centering
		\vspace{-0.5cm}
		\begin{minipage}[t]{0.32\linewidth}
			\centering
			\includegraphics[width=2.2in]{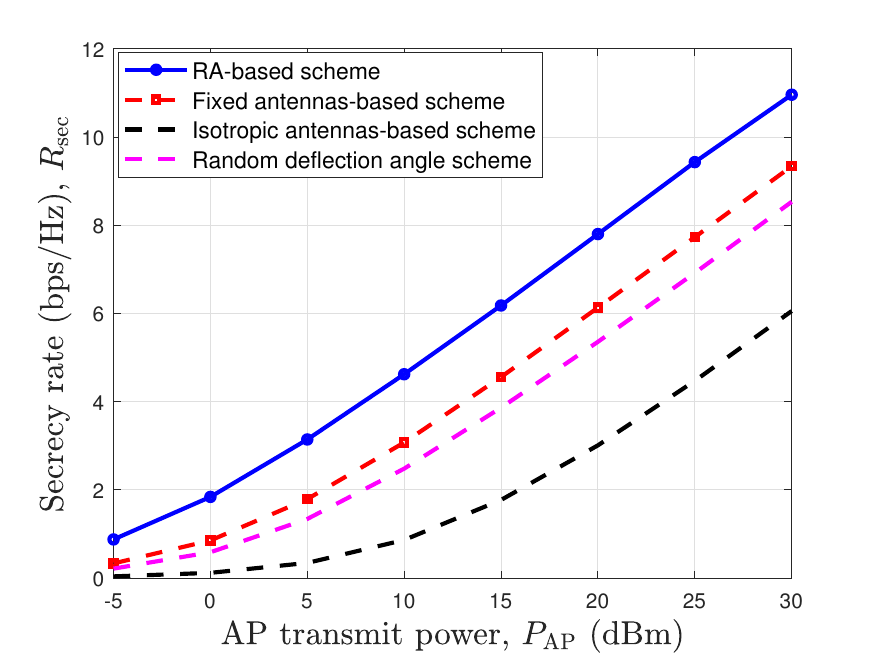}
			\vspace{-0cm}
			\caption{Secrecy rate versus AP transmit power.}
			\label{fig2}
		\end{minipage}%
		\hspace{0.3cm}\begin{minipage}[t]{0.32\linewidth}
			\centering
			\includegraphics[width=2.2in]{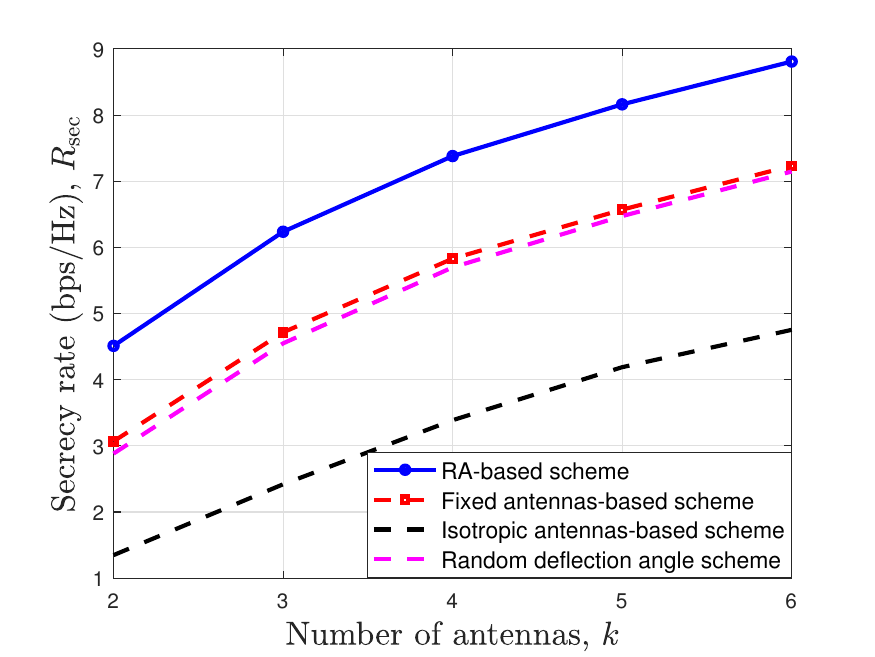}
			\vspace{-0cm}
			\caption{Secrecy rate versus number of antennas ${k}$.}
			\label{fig4}
		\end{minipage}
		\hspace{0.3cm}\begin{minipage}[t]{0.32\linewidth}
			\centering
			\includegraphics[width=2.2in]{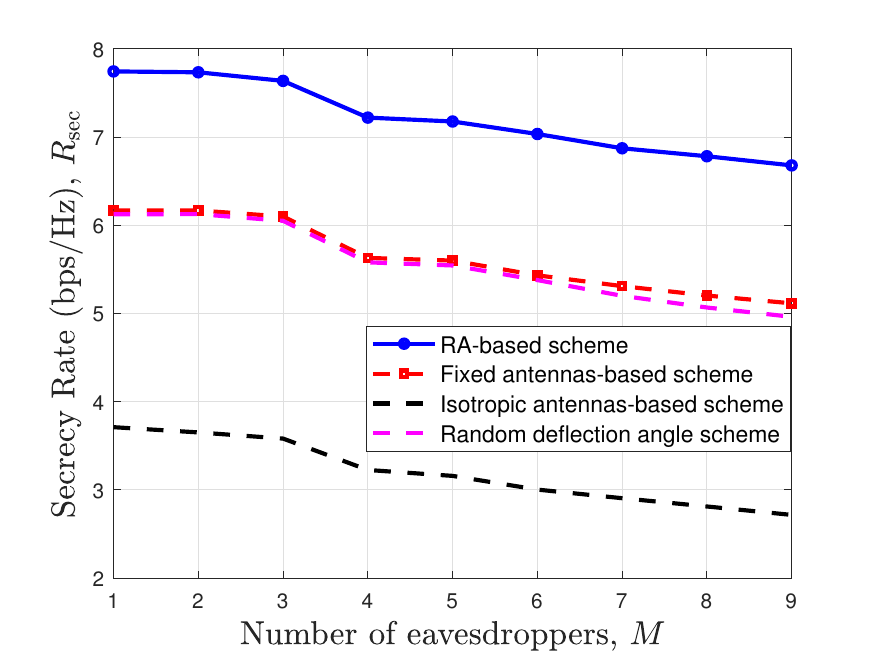}
			\vspace{-0cm}
			\caption{Secrecy rate versus number of eavesdroppers ${M}$.}
			\label{fig5}
		\end{minipage} \vspace{-0.5cm}
	\end{figure*}
	To evaluate the performance of the proposed AO algorithm for optimizing transmit beamforming and RA deflection angles (referred to as the \textbf{RA-based scheme}), we compare it with the following benchmark schemes, all of which adopt the optimal beamforming in (\ref{v_opt}): 
	\begin{itemize}
		\item \textbf{Fixed antennas-based scheme}: In this scheme, the deflection angles are fixed as $\boldsymbol{\Theta} = \boldsymbol{0}_{2 \times K}$.
		\item \textbf{Isotropic antennas-based scheme}: In this scheme, the directional gain is set to $G_0$ = 1 with $p$ = 0 in (\ref{h2}).
		\item \textbf{Random deflection angle scheme}: In this scheme, the deflection angles of each RA, i.e., {$\varphi_{a,k}$} and {$\theta_{z,k}$}, are randomly generated following uniform distributions within $[0,2\pi] $ and $[0,\theta_{\max}]$, respectively.
	\end{itemize}\par
	In Fig. \ref{fig6}, we illustrate the convergence behavior of the proposed AO algorithm. As expected, for all considered different antenna directivity factors $p$, the secrecy rate increases over the iterations and converges within 10 iterations. Additionally, the secrecy rate increases with $p$. This is attributed to the higher directional gain and narrower beamwidth for larger $p$, which is beneficial for antenna array gain pattern reshaping by RAs with radiation power concentration in desired directions.\par
	Fig. \ref{fig7} shows the array gains of different schemes versus angle $\phi \in [0,\pi]$. It is observed that the array gain patterns of all schemes are severely suppressed at the eavesdroppers' directions, indicating the effectiveness of the transmit beamforming design. Although the legitimate user and the eavesdroppers are located in close proximity, the RA-based scheme achieves a higher array gain in the legitimate user's direction by properly adjusting the direction of each RA's main lobe. This result shows that RA can significantly enhance the array gain in the direction of legitimate user while suppressing the radiated power in the directions of the eavesdroppers.\par
	Fig. \ref{fig2} shows the average secrecy rates of different schemes versus the AP’s transmit power, $P_\mathrm{{AP}}$. It is observed that as $P_\mathrm{{AP}}$ increases, the secrecy rates of all schemes increase accordingly. The secrecy rate of the fixed antennas-based scheme is lower than that of the RA-based scheme. This is because the array directional gain pattern of the fixed antenna-based scheme is unchangeable and the radiated power only focuses on a fixed direction. In contrast, by properly optimizing the deflection angles of all RAs, the RA-based scheme can change the array directional gain pattern to focus on a specific region while nulling the signal power in undesired directions, thus achieving a higher secrecy rate than the benchmarks. Notably, since the random deflection angle scheme does not optimize the antenna orientations/boresight directions, it results in suboptimal antenna orientations. Thus, its performance is significantly inferior to that of the RA-based scheme and even worse than that of the fixed antennas-based scheme.  \par
	Fig. \ref{fig4} plots the secrecy rates of different schemes versus the number of antennas. It can be observed that the secrecy rates of the RA-based and other schemes all increase as the number of antennas increases. In particular, the secrecy rate of the RA-based scheme is higher than that of the other schemes, since all RAs effectively leverage the additional DoF to flexibly adjust their 3D orientations/boresight directions for enhancing the achievable rate of the legitimate user.\par 
	Fig. \ref{fig5} shows the secrecy rates obtained by different schemes versus the number of eavesdroppers ${M}$, where the eavesdroppers are randomly and uniformly distributed within the angular range $[\frac{\pi}{12},\frac{11\pi}{12}]$. It is observed that the secrecy rates of all schemes decrease as ${M}$ increases. This is because the collaborative eavesdropping capability of the eavesdroppers grows as $M$ increases, resulting in a degraded secrecy rate. In addition, as $M$ increases, it is more likely for eavesdroppers to appear near the legitimate user, making the beamforming design and antenna orientation adjustment more difficult to simultaneously enhance the directional gain of the legitimate user while suppressing that of the eavesdroppers. Nevertheless, the RA-enabled scheme consistently outperforms benchmark schemes, demonstrating that the proposed algorithm is more robust against the number of eavesdroppers.

	 \section{Conclusion}
	 In this letter, we proposed a novel RA-enabled secure communication system, where the deflection angles of RAs can be adjusted to change the array directional gain pattern of the RA-based array to improve the secrecy rate. To obtain a high-quality solution for the deflection angles, a low-complexity AO algorithm was developed to jointly optimize the transmit beamforming weights and deflection angles of all RAs at the AP. Simulation results demonstrated that by adjusting the orientation/boresight direction of each antenna, the RA-based array significantly enhances the array gain towards the legitimate user while suppressing the power towards the eavesdroppers, thereby yielding a higher secrecy rate as compared to various benchmark schemes.

\bibliography{document}

\begin{thebibliography}{10}
\providecommand{\url}[1]{#1}
\csname url@samestyle\endcsname
\providecommand{\newblock}{\relax}
\providecommand{\bibinfo}[2]{#2}
\providecommand{\BIBentrySTDinterwordspacing}{\spaceskip=0pt\relax}
\providecommand{\BIBentryALTinterwordstretchfactor}{4}
\providecommand{\BIBentryALTinterwordspacing}{\spaceskip=\fontdimen2\font plus
\BIBentryALTinterwordstretchfactor\fontdimen3\font minus
  \fontdimen4\font\relax}
\providecommand{\BIBforeignlanguage}[2]{{%
\expandafter\ifx\csname l@#1\endcsname\relax
\typeout{** WARNING: IEEEtran.bst: No hyphenation pattern has been}%
\typeout{** loaded for the language `#1'. Using the pattern for}%
\typeout{** the default language instead.}%
\else
\language=\csname l@#1\endcsname
\fi
#2}}
\providecommand{\BIBdecl}{\relax}
\BIBdecl

\bibitem{01}
K.~{Shim}, ``A survey of public-key cryptographic primitives in wireless sensor
  networks,'' \emph{IEEE Commun. Surveys Tuts.}, vol.~18, no.~1, pp. 577--601,
  1st Quart. 2016.

\bibitem{0}
X.~{Chen}, D.~W.~K. {Ng}, W.~H. {Gerstacker}, and H.-H. {Chen}, ``{A} survey on
  multiple-antenna techniques for physical layer security,'' \emph{IEEE Commun.
  Surveys Tuts.}, vol.~19, no.~2, pp. 1027--1053, 2nd Quart. 2017.

\bibitem{1}
H.~{Xing}, L.~{Liu}, and R.~{Zhang}, ``{Secrecy wireless information and power
  transfer in fading wiretap channel},'' \emph{IEEE Trans. Veh. Technol.},
  vol.~65, no.~1, pp. 180--190, Jul. 2016.

\bibitem{2}
R.~{Negi} and S.~{Goel}, ``{Secret communication using artificial noise},'' in
  \emph{Proc. IEEE Vehicular Tech. Conf.}, vol.~62, no.~3, Sept. 2005, pp.
  1906--1910.

\bibitem{3}
G.~{Zheng}, L.-C. {Choo}, and K.-K. {Wong}, ``{Optimal cooperative jamming to
  enhance physical layer security using relays},'' \emph{IEEE Trans. Signal
  Process.}, vol.~59, no.~3, pp. 1317--1322, Mar. 2011.

\bibitem{4}
A.~{Khisti} and G.~W. {Wornell}, ``{Secure transmission with multiple antennas
  I: {The} {MISOME} wiretap channel},'' \emph{IEEE Trans. Inf. Theory},
  vol.~56, no.~7, pp. 3088--3104, Jul. 2010.

\bibitem{5}
L.~{Liu}, R.~{Zhang}, and K.~{Chua}, ``{Secrecy wireless information and power
  transfer with {MISO} beamforming},'' \emph{IEEE Trans. Signal Process.},
  vol.~62, no.~7, pp. 1850--1863, Apr. 2014.

\bibitem{6}
J.~{Zhu}, R.~{Schober}, and V.~K. {Bhargava}, ``{Linear} precoding of data and
  artificial noise in secure massive {MIMO} systems,'' \emph{IEEE Trans.
  Wireless Commun.}, vol.~15, no.~3, pp. 2245--2261, Mar. 2016.

\bibitem{7}
Q.~{Wu}, B.~{Zheng}, T.~{Ma}, and R.~{Zhang}, ``{Modeling and optimization for
  rotatable antenna enabled wireless communication},'' \emph{arXiv preprint
  arXiv:2411.08411}, 2024.

\bibitem{8}
B.~{Zheng}, Q.~{Wu}, and R.~{Zhang}, ``{Rotatable antenna enabled wireless
  communication: {Modeling} and optimization},'' \emph{arXiv preprint
  arXiv:2501.02595}, 2025.

\bibitem{11}
X.~{Shao}, Q.~{Jiang}, and R.~{Zhang}, ``{6D movable antenna based on user
  distribution: Modeling and optimization},'' \emph{IEEE Trans. Wireless
  Commun.}, vol.~24, no.~1, pp. 355--370, Jan. 2025.

\bibitem{9}
C.~{Zhou}, C.~{You}, B.~{Zheng}, X.~{Shao}, and R.~{Zhang}, ``{Rotatable}
  antennas for integrated sensing and communications,'' \emph{arXiv preprint
  arXiv:2503.10472}, 2025.

\bibitem{10}
Y.~{Xie}, W.~{Mei}, D.~{Wang}, B.~{Ning}, Z.~{Chen}, J.~{Fang}, and W.~{Guo},
  ``{THz} beam squint mitigation via {3D} rotatable antennas,'' \emph{arXiv
  preprint arXiv:2503.08134}, 2025.

\end{thebibliography}
\end{CJK}
\end{document}